# PACS: Prediction and analysis of cancer subtypes from multi-omics data based on a multi-head attention mechanism model

Liangrui Pan
College of Computer Science and Electronic Engineering
HunanUniversity
Chang Sha, China
panlr@ hnu.edu.cn

Dazheng Liu
College of Computer Science and Electronic Engineering
HunanUniversity
Chang Sha, China
liudz@hnu.edu.cn

Zhichao Feng
Department of Radiology
Third Xiangya Hospital
Central South University
Chang Sha, China
fengzc2016@163.com

Wenjuan Liu*
College of Computer Science and Electronic Engineering
Hunan University
Chang Sha, China
liuwenjuan89@hnu.edu

Shaoliang Peng*
College of Computer Science and Electronic Engineering
Hunan University
Chang Sha, China
slpeng@hnu.edu.cn

*Abstract*—Due to the high heterogeneity and clinical characteristics of cancer, there are significant differences in multi-omic data and clinical characteristics among different cancer subtypes. Therefore, accurate classification of cancer subtypes can help doctors choose the most appropriate treatment options, improve treatment outcomes, and provide more accurate patient survival predictions. In this study, we propose a supervised multi-head attention mechanism model (SMA) to classify cancer subtypes successfully. The attention mechanism and feature sharing module of the SMA model can successfully learn the global and local feature information of multi-omics data. Second, it enriches the parameters of the model by deeply fusing multi-head attention encoders from Siamese through the fusion module. Validated by extensive experiments, the SMA model achieves the highest accuracy, F1 macroscopic, F1 weighted, and accurate classification of cancer subtypes in simulated, single-cell, and cancer multi-omics datasets compared to AE, CNN, and GNN-based models. Therefore, we contribute to future research on multi-omics data using our attention-based approach.

## I. INTRODUCTION

With the development of third-generation high-throughput sequencing technologies, it has become efficient and fast to sequence and analyze the genomes, transcriptomes, and DNA/RNA samples of organisms, thereby generating a large amount of multi-omics data [1], [2]. Therefore, the development of high-throughput sequencing technology is closely related to the advancement of multi-omics research. The wealth of genomic, transcriptomic, and epigenomic data obtained through these technologies has enriched the scope of multi-omics studies. These data allow for a comprehensive understanding of molecular changes at different levels within an organism, revealing the relationship between genes and phenotypes and uncovering the underlying mechanisms of biological processes. Furthermore, the development of multi-omics research has also propelled advancements and applications of high-throughput sequencing technologies. In summary, the development of high-throughput sequencing and multi-omics research is mutually beneficial, enabling comprehensive analyses of cancer development and treatment.

The development of multi-omics integrates high-throughput technologies and data analysis methods. Multi-omics data include genomics, transcriptomics, and proteomics [3]–[5]. Genomics research commonly involves comprehensive sequencing and analysis of an organism's genetic material, such as DNA. The rapid progress in single-cell genomics allows us to study genetic variations and evolutionary processes within individual cells. Transcriptomics focuses on the entirety of an organism's transcriptional products (RNA) during specific periods, in specific tissues, or under certain environmental conditions [6]. Through high-throughput transcriptomic sequencing technologies, we can understand the spectrum of gene expression, identify differentially expressed genes, alternative splicing events, and reveal the structure and function of gene regulatory networks. Proteomics research focuses on the expression levels, modification states, and interactions of proteins within an organism. The development of high-throughput mass spectrometry technology enables large-scale and efficient protein identification and quantitative analysis, further unveiling protein functions and compositions. Therefore, multi-omics plays an increasingly important role in life sciences and medicine.

Given the heterogeneity and individual differences in cancer, doctors need to classify cancer into subtypes to provide more accurate guidance in treatment and prognosis assessment [7]. This helps realize the concept of precision medicine, providing patients with more individualized and targeted treatment plans and improving treatment efficacy and patient quality of life. Some existing machine learning methods utilize multi-omics data to classify cancer subtypes and have achieved high classification accuracy. Similarity Network Fusion (SNF) integrates different omics data to enhance understanding of tumor development [8]. It primarily uses the Euclidean distance to measure patient similarity [8]. As an extension of SNF, Deep Subspace Fusion Clustering (DSFC) employs autoencoders and data self-expression techniques to guide deep subspace models, effectively expressing discriminative similarities among patients [9]. Several supervised and semi-supervised deep learning methods can accurately classify breast cancer subtypes by identifying nonlinear relationships in multi-omics data [10]. The neighbourhood-based Multi-omics Clustering Method (NEMO) extensively tested ten cancer datasets and achieved the best subtype classification performance in benchmark experiments [11]. Furthermore, based on DNA methylation data, they proposed a two-stage XAI-methylation biomarker discovery framework—an explainable AI-based biological biomarker discovery framework applied to DNA methylation data to identify a subset of biomarkers for breast cancer classification [12].

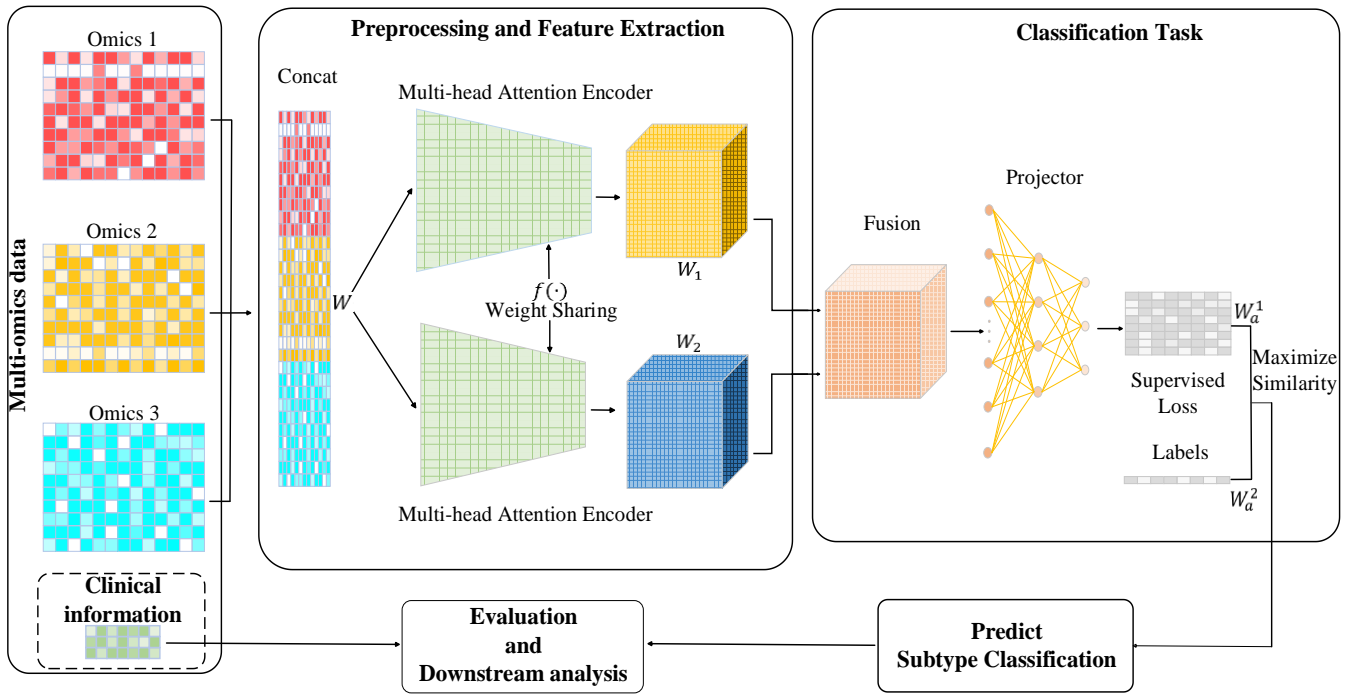

Fig. 1. Flowchart of the SMA model for predicting cancer subtypes.

Deep learning to analyze multi-omics data to classify cancer subtypes is still mainstream [13], [14]. The deep learning model has robust learning ability and can automatically learn advanced feature representations through a large amount of data. After sufficient training, deep learning models can perform well on large-scale data sets and have good generalization ability. However, the deep learning method requires a large amount of data, takes a long time to train, and needs better interpretability, making it challenging to perform correlation analysis on multi-omics data.

Recently, inspired by the attention mechanism, it can focus on features that play a key role in classification results and suppress noisy or irrelevant features in terms of feature selection and importance evaluation [14], [15]. There are also complex interrelationships and dependencies between multi-omics data. The attention mechanism can help the model capture contextual information to understand their relationship better. The attention mechanism can extract more richer, more meaningful representations through the weighted combination of features. Second, the attention mechanism can enhance the robustness and generalization ability of the model. To a certain extent, it can explain the focus and decision dependence of the model in the classification process. This paper proposes an attention-based multi-omics cancer subtype classification model to predict cancer subtypes. The contributions of this paper are as follows:

1. We propose an attention-based model to learn the features of multi-omics data, which can capture both global and local features by perceiving the context.
2. The SMA model consists of a twin attention mechanism extraction module. The twin modules share the results of feature extraction and perform supervised classification through feature fusion.
3. We conducted experiments on simulated, single-cell, and cancer multi-omics datasets. The results demonstrate that the SMA model can accurately classify cancer subtypes.

II. METHODS

A. Multi-Head Attention Encoder

As shown in Fig.1, first, Multi-omics data must be preprocessed, including data cleaning, alignment of sample features, and sample screening. The experiment selects some representative features as training data in different data sets. Then it normalizes the data of various omics to ensure that data of different scales can be compared and merged. These steps are mainly to improve the quality of the data, improve the accuracy of the model, reduce the complexity of the model calculation, and speed up the subsequent analysis.

After the multi-omics data are preprocessed, they are spliced according to the order of the samples. Features between different omics are added together. The matrix after feature splicing will be fed into two encoders based on a multi-head attention mechanism. The encoder of the multi-head attention mechanism mainly includes a position encoding module, an attention mechanism module, a feature forward transmission module and a perceptron module [16]. First, the position encoding module linearly maps the matrix to features to generate cls tokens and small feature tokens. The process is:

$$feature = input\ embedding + positional\ encoding \quad (1)$$

Position encoding adds position information to each position of the input sequence, generally using a combination of sine and cosine functions. The feature linear mapping will then be fed into the encoder of the multi-head attention mechanism for feature extraction. It copies features into the query, key, and value feature space, and each is split into multiple heads for processing. Calculate the attention score (i.e., the similarity between Q and K) for each attention head and normalize the attention score to a probability distribution [17]. Its process can be described as:

$$Attention(Q_i, K_i, V_i) = soft\max(Q_i * K_i)/sqrt(d_k) * V_i \quad (2)$$

The Softmax function normalizes the attention score to a probability distribution, $K$ is the feature dimension of $d_k$. Then the output of each attention head is spliced, and projected and integrated through linear transformation, as follows:

$$MultiHead = Concatenate(Attention(Q_i, K_i, V_i)) * W \quad (3)$$

Where concatenate is to splice the output of each attention head and is the feature weight of learning. Then, after extracting features with attention, the feature matrix is passed to the feature forward transfer module and the perceptron module. They are mainly to enhance the representation ability and generalization ability of the model. The feature forward transfer module comprises two fully connected layers and an activation function. Multilayer Perceptron is composed of multiple fully connected layers and activation functions [18]. Stacking multiple hidden layers can achieve higher-dimensional and more complex feature transformations.

*B. Supervised training strategy*

The symmetric multi-head attention mechanism encoder extracts feature from multiple sets of learning data and generates feature matrices $W_1$ and $W_2$. We use a feature fusion method of element-wise multiplication to fuse the features of $W_1$ and $W_2$, resulting in a fused feature vector. This method can highlight the unique features of the encoder and improve the performance and generalization ability of the model.

The fused feature matrix is fed into a three-layer perceptron for normalization, further projected into a new feature space, and a new feature matrix $W_a^1$ is generated. We use the cross-entropy loss function to calculate the error between a single predicted sample and its corresponding label, with the process as follows [17]:

$$L_i = -[y\log \hat{y} + (1-y)\log(1-\hat{y})] \quad (4)$$

When we calculate the distance between the feature matrix $W_a^1$ and the label, the overall loss function can be expressed as:

$$L = -\sum_{i=1}^{N}[y_i \log \hat{y}_i + (1-y_i)\log(1-\hat{y}_i)] \quad (5)$$

## III. EXPERIMENTS

*A. Materials*

**Simulated Dataset:** This dataset is generated by the InterSIM CRAN package and consists of complex and interrelated multi-omics data [19]. It includes DNA methylation, mRNA gene expression, and protein expression data from 100 samples, with clusters set to 5, 10, or 15. Furthermore, the software generates clusters for each sample under two conditions: "equal" and "heterogeneous" [20]. All clusters have the same size under the "equal" condition, while the cluster sizes are randomly variable under the "heterogeneous" condition. The simulated dataset is similar to a real multi-omics dataset, where the sample proportions in each cluster can be the same or different [20].
**Single-cell dataset**: This dataset includes 206 single-cell samples from three cancer cell lines (HTC, Hela, and K562). Two types of omics data were obtained, namely single-cell chromatin accessibility and single-cell gene expression data. The features of these two types of omics data are 49,073 and 207,203, respectively [20]–[22].
**Cancer Multi-Omics Dataset**: This dataset is derived from the cancer multi-omics dataset in The Cancer Genome Atlas (TCGA) and consists of gene expression, DNA methylation, and miRNA expression data [23]. The dataset includes breast cancer (BRCA), glioblastoma (GBM), sarcoma (SARC), lung adenocarcinoma (LUAD), and stomach cancer (STAD) from TCGA. Other cancer types are selected from the baseline dataset, including colon cancer (Colon), acute myeloid leukemia (AML), kidney cancer (Kidney), melanoma, and ovarian cancer. Datasets can be accessed at http://acgt.cs.tau.ac.il/multi_omic_benchmark/download.html [8], [20], [24].

*B. Experimental details*

The SMA model was developed using Python 3.8.5 on the PyTorch 1.7.1 platform. The training of the model was performed using an NVIDIA RTX4090 GPU. In order to optimize the performance of the model, five main hyperparameters were adjusted: batch size, epochs, optimizer, learning rate, and weight decay. Finally, the batch size was set to 128, epochs were set to 500, the learning rate was set to 3e-3, and the weight decay was set to 0.01.

*C. Evaluation*

In classification tasks, we choose Accuracy, F1 macro, and F1 weighted to evaluate the performance of the SMA model [13]. Accuracy refers to the proportion of correct classifications in all samples by the SMA model. F1 macro refers to the average F1 value of all categories, where F1 is the harmonic mean of precision and recall. F1 weighted refers to the weighted average of F1 values calculated according to the proportion of samples in all categories. The larger the values of Accuracy, F1 macro, and F1 weighted, the better the performance of the model [13].

## IV. RESULT AND DISCUSSION

*A. Evaluation of the SMA model on simulated datasets classification tasks*

We compared the classification performance of SMA with six common methods for classifying six types of omics data: (1) lfNN model: each omics vector is concatenated into a feature vector as the input of the model, multiple neural networks are used for feature extraction, and Softmax is used as the last layer for output classification [13]. (2) efNN model: each omics vector is used as the input of the model, multiple neural networks are used for feature extraction, and the outputs are connected into a vector and used Softmax as the last layer for output classification. (3) lfCNN model: it is similar to efNN, but with added convolutional and pooling layers. Multiple omics vectors are concatenated into a feature vector, which is sent to the convolutional and pooling layers, and the output features are flattened and sent to the fully connected network for final prediction. (4) efCNN model [13]: it is similar to lfNN. Each omics vector is sent to the convolutional and pooling layers, the output features are flattened, connected, and fed into a fully connected neural network for final prediction. (5) moGCN model: it uses GCN to learn the features of omics data and perform classification tasks. To perform omics-specific classification, a multi-layer GCN needs to be built for each type of omics data. (6) moGAT model: GAT in moGCN is replaced by moGAT model. In the testing method, lfNN, efNN, lfCNN, efCNN, moGCN, moGAT, and SMA models are trained by directly concatenating preprocessed multiple omics data as input, and all models use the same preprocessed data for training [13]. The classification performance of all models can be referred to the data in TABLE I under the conditions of equal and heterogeneous in the simulated dataset.

TABLE I. PERFORMANCE OF SEVEN SUPERVISED METHODS IN THE CONDITION THAT ALL CLUSTERS HAVE THE SAME SIZE.

| Methods | 5 clusters of random sizes | | | 10 clusters of random sizes | | | 15 clusters of random sizes | | |
|---|---|---|---|---|---|---|---|---|---|
| | Accuracy | F1 macro | F1 weighted | Accuracy | F1 macro | F1 weighted | Accuracy | F1 macro | F1 weighted |
| lfNN | **1.0** | **1.0** | **1.0** | 0.900 | 0.825 | 0.871 | 0.860 | 0.716 | 0.817 |
| efNN | **1.0** | **1.0** | **1.0** | **1.0** | **1.0** | **1.0** | **1.0** | **1.0** | **1.0** |
| lfCNN | **1.0** | **1.0** | **1.0** | 0.480 | 0.454 | 0.361 | 0.880 | 0.758 | 0.840 |
| efCNN | **1.0** | **1.0** | **1.0** | **1.0** | **1.0** | **1.0** | 0.760 | 0.572 | 0.670 |
| moGCN | **1.0** | **1.0** | **1.0** | **1.0** | **1.0** | **1.0** | **1.0** | **1.0** | **1.0** |
| moGAT | **1.0** | **1.0** | **1.0** | **1.0** | **1.0** | **1.0** | **1.0** | **1.0** | **1.0** |
| SMA | **1.0** | **1.0** | **1.0** | **1.0** | **1.0** | **1.0** | **1.0** | **1.0** | **1.0** |

The experiment selected samples from 5 clusters of random sizes, 10 clusters of random sizes, and 15 clusters of random sizes [13]. These seven supervised methods were essentially designed for sample classification, and they classified samples of true clustering (subtypes). In the classification task, in order to quantitatively evaluate the methods of the seven supervised models, we used a simple random cross-validation method to train and test the models. At the same time, all models applied three evaluation indicators, Accuracy, F1 macro, and F1 weighted, to measure the performance of classification. From TABLE I, the efNN, moGCN, moGAT, and SMA models performed well in the multi-omics classification task. The efCNN model performed significantly lower than the other six models in the sample classification task of 15 clusters of random sizes. The lfNN model performed significantly lower than the other six models in the sample classification task of 10 clusters of random sizes. This may be because the simulated dataset experienced overfitting after multiple layers of convolutional layers and pooling layers, leading to misclassification in the model. The lfNN model only achieved the best performance in the sample classification task of 5 clusters of random sizes. This may be because the lfNN model failed to learn multi-omics features in the process of feature extraction, leading to misclassification.

*B. Evaluation of the SMA model in single-cell data classification tasks*

In classification tasks, similar to the evaluation methods of lfNN, efNN, lfCNN, efCNN, moGCN, moGAT and SMA models on simulated datasets, we will investigate the performance of these models on single-cell datasets [13]. All models use simple cross-validation methods to classify samples of three cancer cell lines, and the performance of classification is measured by three evaluation metrics: Accuracy, F1 macro, and F1 weighted.

As described in Table II, we found that the lfNN, efNN, moGCN, moGAT, and SMA models all reached their peak performance in terms of Accuracy, F1 macro, and F1 weighted evaluation metrics, indicating that these models achieved the best performance in the classification task. However, the lfCNN and efCNN models still did not perform as well as other models in testing, partly because the number of convolutional and pooling layers was relatively small, and the effect of feature extraction was not significant enough. Another reason could be the lack of regularization penalties in the model or other reasons, such as inappropriate learning rate or batch size, etc.

TABLE II. PERFORMANCE OF SIX SUPERVISED METHODS ON SINGLE-CELL MULTI-OMICS DATASETS

| | Accuracy | F1 macro | F1 weighted |
|---|---|---|---|
| lfNN | **1.0** | **1.0** | **1.0** |
| efNN | **1.0** | **1.0** | **1.0** |
| lfCNN | 0.962 | 0.952 | 0.962 |
| efCNN | 0.981 | 0.981 | 0.981 |
| moGCN | **1.0** | **1.0** | **1.0** |
| moGAT | **1.0** | **1.0** | **1.0** |
| SMA | **1.0** | **1.0** | **1.0** |

*C. Evaluation of SMA Model on Cancer Dataset classification tasks*

In the classification task, similar to the methods used to evaluate models on simulated and single-cell datasets for lfNN, efNN, lfCNN, efCNN, moGCN, moGAT and SMA models, experiments were selected on five datasets with real cancer subtypes experiment [13]. These methods classify real samples of cancer subtypes. All model training and testing use a simple cross-validation method, and measure the classification performance through three evaluation indicators: Accuracy, F1 macro, and F1 weighted. For each cancer dataset, we selected three omics data samples, obtaining 59, 272, 206, 144, and 198 samples for BRCA, GBM, SARC, LUAD, and STAD, respectively 30. BRCA includes five cancer subtypes, namely LuminalA, LuminalB, Basal-like, Normal-like, and HER2-enriched [25]. GBM includes 4 cancer subtypes, namely Proneural, Classical, Mesenchymal, and Neural [26]. SARC also includes five cancer subtypes, dedifferentiated liposarcoma, leiomyosarcoma, undifferentiated pleomorphic sarcoma, myxofibrosarcoma, malignant peripheral nerve sheath tumor, synovial sarcoma [27], [28]. LUAD includes four cancer subtypes, namely formerly bronchioid, formerly squamoid, and formerly magnoid [29]. STAD includes Epstein–Barr virus, microsatellite instability, genomically stable, chromosomal instability [30].

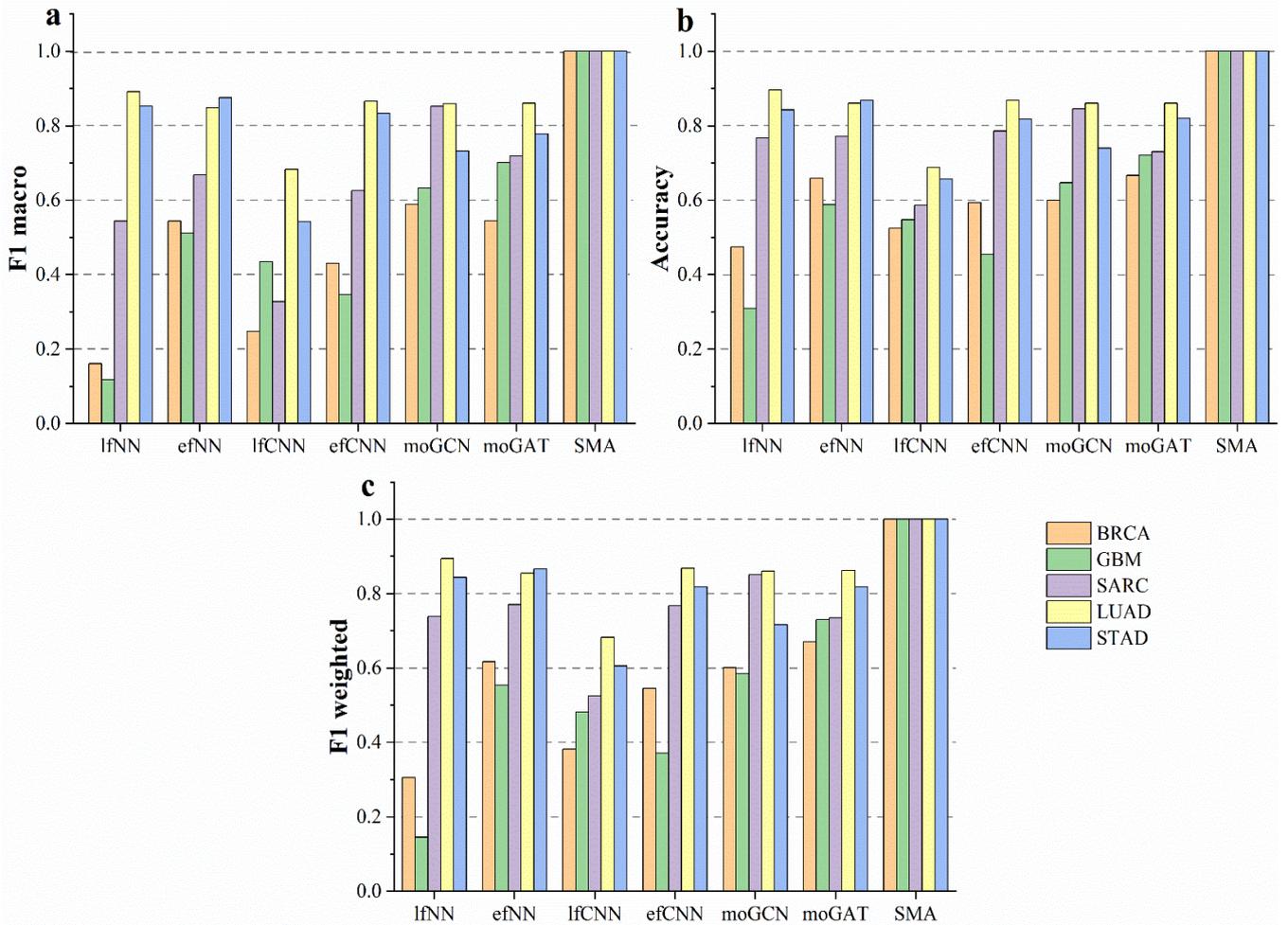

Fig. 2. Performance of seven supervised methods using cancer benchmark datasets in classification tasks.

As shown in Fig. 2, the SMA model achieved high performance in accurately classifying cancer subtypes in BRCA, GBM, SARC, and STAD, with all three evaluation metrics (Accuracy, F1 macro, and F1 weighted) reaching 1. However, when classifying subtypes in LUAD, the SMA model achieved only high performance with respective scores of 0.958, 0.93, and 0.91. Compared to other models, the SMA model outperforms them in classification tasks. This may be because the SMA model can simultaneously pay attention to the positional information of the input sequence, thus capturing global information. Additionally, the SMA model has a deeper structure and more parameters, which can help learn more features and improve classification accuracy. Therefore, the SMA model can serve as a standard method for classifying cancer multi-omics data.

*D. Discussion*

The rapid development of high-throughput sequencing technology has made it possible to use molecular-level data to personalize medicine in unprecedented detail [31]. Multi-omics technologies can integrate different types of data to form more comprehensive and integrated datasets, which can better reveal the complexity of biological systems. This paper proposes a SMA model to classify cancer subtypes successfully. The experimental system analyzes three representative cancer multi-omic datasets (simulated, single-cell, and cancer multi-omic datasets) in three different contexts.

The SMA model performed well in classification tasks when evaluated on simulated multi-omics data. Compared to efNN, moGCN, and moGAT models, the SMA model had higher feature extraction efficiency and better robustness. This may be because the attention mechanism can capture long-range dependencies in each omics data and perform parallel computation on the entire sequence through position encoding, avoiding information loss and gradient vanishing. The SMA model also exhibited high diagnostic performance when evaluated on single-cell datasets. However, the diagnostic performance of efCNN and lfCNN models based on CNN architecture was poor.

This may be because CNN loses more feature details in one-dimensional feature extraction, resulting in poor diagnostic performance of the model. When evaluated on cancer multi-omics datasets, the SMA model achieved the best performance in subtype classification of cancer in BRCA, GBM, SARC, and STAD. However, the SMA model had a misdiagnosis in classifying LUAD. Therefore, we believe the SMA model is a standard solution for multi-omics data classification.

## V. CONCLUSION

In this study, we propose a supervised SMA model to classify cancer subtypes successfully. The attention mechanism and feature sharing module of the SMA model can successfully learn the global and local feature information of multi-omics data. Second, it enriches the parameters of the model by deeply fusing the multi-head attention mechanism encoder from Siamese through the fusion module. Through extensive experimental verification, we found that the attention mechanism can obtain rich information from the context of multi-omics data. Compared to AE, CNN, and GNN-based models, the SMA model achieved the highest accuracy, F1 macroscopic, F1 weighted, and accurately classified cancer subtypes in simulated, single cell, and cancer multi-omics datasets. Therefore, we contribute to future research on multi-omics data using our attention-based approach.

## ACKNOWLEDGEMENTS

This work was supported by NSFC Grants U19A2067; National Key R&D Program of China 2022YFC3400400; Top 10 Technical Key Project in Hunan Province 2023GK1010, Key Technologies R&D Program of Guangdong Province (2023B1111030004 to FFH). The Funds of State Key Laboratory of Chemo/Biosensing and Chemometrics, the National Supercomputing Center in Changsha (http://nscc.hnu.edu.cn/), and Peng Cheng Lab.

## REFERENCE

[1] J. A. Reuter, D. V. Spacek, and M. P. Snyder, "High-Throughput Sequencing Technologies," *Mol. Cell*, vol. 58, no. 4, pp. 586–597, May 2015, doi: 10.1016/j.molcel.2015.05.004.
[2] H. Satam et al., "Next-Generation Sequencing Technology: Current Trends and Advancements," *Biology*, vol. 12, no. 7, 2023, doi: 10.3390/biology12070997.
[3] D. B. Seal, V. Das, S. Goswami, and R. K. De, "Estimating gene expression from DNA methylation and copy number variation: A deep learning regression model for multi-omics integration," *Genomics*, vol. 112, no. 4, pp. 2833–2841, Jul. 2020, doi: 10.1016/j.ygeno.2020.03.021.
[4] D. R. Mani et al., "Cancer proteogenomics: current impact and future prospects," *Nat. Rev. Cancer*, vol. 22, no. 5, pp. 298–313, May 2022, doi: 10.1038/s41568-022-00446-5.
[5] F. Nassiri et al., "A clinically applicable integrative molecular classification of meningiomas," *Nature*, vol. 597, no. 7874, pp. 119–125, Sep. 2021, doi: 10.1038/s41586-021-03850-3.
[6] R. Lowe, N. Shirley, M. Bleackley, S. Dolan, and T. Shafee, "Transcriptomics technologies," *PLoS Comput. Biol.*, vol. 13, no. 5, p. e1005457, 2017.
[7] O. Yersal and S. Barutca, "Biological subtypes of breast cancer: Prognostic and therapeutic implications," *World J. Clin. Oncol.*, vol. 5, no. 3, p. 412, 2014.
[8] B. Wang et al., "Similarity network fusion for aggregating data types on a genomic scale," *Nat. Methods*, vol. 11, no. 3, pp. 333–337, Mar. 2014, doi: 10.1038/nmeth.2810.
[9] B. Yang, Y. Zhang, S. Pang, X. Shang, X. Zhao, and M. Han, "Integrating Multi-Omic Data With Deep Subspace Fusion Clustering for Cancer Subtype Prediction," *IEEE/ACM Trans. Comput. Biol. Bioinform.*, vol. 18, no. 1, pp. 216–226, Feb. 2021, doi: 10.1109/TCBB.2019.2951413.
[10] F. Cristovao et al., "Investigating Deep Learning Based Breast Cancer Subtyping Using Pan-Cancer and Multi-Omic Data," *IEEE/ACM Trans. Comput. Biol. Bioinform.*, vol. 19, no. 1, pp. 121–134, Feb. 2022, doi: 10.1109/TCBB.2020.3042309.
[11] N. Rappoport and R. Shamir, "NEMO: cancer subtyping by integration of partial multi-omic data," *Bioinformatics*, vol. 35, no. 18, pp. 3348–3356, 2019.
[12] S. Rajpal et al., "XAI-MethylMarker: Explainable AI approach for biomarker discovery for breast cancer subtype classification using methylation data," *Expert Syst. Appl.*, vol. 225, p. 120130, Sep. 2023, doi: 10.1016/j.eswa.2023.120130.
[13] D. Leng et al., "A benchmark study of deep learning-based multi-omics data fusion methods for cancer," *Genome Biol.*, vol. 23, no. 1, p. 171, Aug. 2022, doi: 10.1186/s13059-022-02739-2.
[14] P. Gong et al., "Multi-omics integration method based on attention deep learning network for biomedical data classification," *Comput. Methods Programs Biomed.*, vol. 231, p. 107377, Apr. 2023, doi: 10.1016/j.cmpb.2023.107377.
[15] D. Ouyang et al., "Integration of multi-omics data using adaptive graph learning and attention mechanism for patient classification and biomarker identification," *Comput. Biol. Med.*, p. 107303, Aug. 2023, doi: 10.1016/j.compbiomed.2023.107303.
[16] K. Han et al., "A Survey on Vision Transformer," *IEEE Trans. Pattern Anal. Mach. Intell.*, pp. 1–1, 2022, doi: 10.1109/TPAMI.2022.3152247.
[17] L. Pan et al., "Noise-reducing attention cross fusion learning transformer for histological image classification of osteosarcoma," *Biomed. Signal Process. Control*, vol. 77, p. 103824, Aug. 2022, doi: 10.1016/j.bspc.2022.103824.
[18] A. Dosovitskiy et al., "An image is worth 16x16 words: Transformers for image recognition at scale," *ArXiv Prepr. ArXiv201011929*, 2020.
[19] P. Chalise, R. Raghavan, and B. L. Fridley, "InterSIM: Simulation tool for multiple integrative 'omic datasets,'" *Comput. Methods Programs Biomed.*, vol. 128, pp. 69–74, May 2016, doi: 10.1016/j.cmpb.2016.02.011.
[20] L. Pan et al., "Multi-Head Attention Mechanism Learning for Cancer New Subtypes and Treatment Based on Cancer Multi-Omics Data," *ArXiv Prepr. ArXiv230704075*, 2023.
[21] L. Liu et al., "Deconvolution of single-cell multi-omics layers reveals regulatory heterogeneity," *Nat. Commun.*, vol. 10, no. 1, p. 470, Jan. 2019, doi: 10.1038/s41467-018-08205-7.
[22] J. Lee, D. Y. Hyeon, and D. Hwang, "Single-cell multiomics: technologies and data analysis methods," *Exp. Mol. Med.*, vol. 52, no. 9, pp. 1428–1442, Sep. 2020, doi: 10.1038/s12276-020-0420-2.
[23] K. Tomczak, P. Czerwińska, and M. Wiznerowicz, "ReviewThe Cancer Genome Atlas (TCGA): an immeasurable source of knowledge," *Contemp. Oncol. Onkol.*, pp. 68–77, 2015, doi: 10.5114/wo.2014.47136.
[24] E. F. Franco et al., "Performance Comparison of Deep Learning Autoencoders for Cancer Subtype Detection Using Multi-Omics Data," *Cancers*, vol. 13, no. 9, 2021, doi: 10.3390/cancers13092013.
[25] A. Llombart-Cussac et al., "HER2-enriched subtype as a predictor of pathological complete response following trastuzumab and lapatinib without chemotherapy in early-stage HER2-positive breast cancer (PAMELA): an open-label, single-group, multicentre, phase 2 trial," *Lancet Oncol.*, vol. 18, no. 4, pp. 545–554, Apr. 2017, doi: 10.1016/S1470-2045(17)30021-9.
[26] R. G. W. Verhaak et al., "Integrated Genomic Analysis Identifies Clinically Relevant Subtypes of Glioblastoma Characterized by Abnormalities in PDGFRA, IDH1, EGFR, and NF1," *Cancer Cell*, vol. 17, no. 1, pp. 98–110, Jan. 2010, doi: 10.1016/j.ccr.2009.12.020.
[27] S. Wei, E. Henderson-Jackson, X. Qian, and M. M. Bui, "Soft Tissue Tumor Immunohistochemistry Update: Illustrative Examples of Diagnostic Pearls to Avoid Pitfalls," *Arch. Pathol. Lab. Med.*, vol. 141, no. 8, pp. 1072–1091, Aug. 2017, doi: 10.5858/arpa.2016-0417-RA.
[28] B. C. Widemann and A. Italiano, "Biology and Management of Undifferentiated Pleomorphic Sarcoma, Myxofibrosarcoma, and Malignant Peripheral Nerve Sheath Tumors: State of the Art and Perspectives," *J. Clin. Oncol.*, vol. 36, no. 2, pp. 160–167, Jan. 2018, doi: 10.1200/JCO.2017.75.3467.

[29] E. A. Collisson *et al.*, "Comprehensive molecular profiling of lung adenocarcinoma," *Nature*, vol. 511, no. 7511, pp. 543–550, Jul. 2014, doi: 10.1038/nature13385.

[30] S. Derks *et al.*, "Abundant PD-L1 expression in Epstein-Barr Virus-infected gastric cancers," *Oncotarget Vol 7 No 22*, 2016, Accessed: Jan. 01, 2016. [Online]. Available: https://www.oncotarget.com/article/9076/text/

[31] T. Wang *et al.*, "MOGONET integrates multi-omics data using graph convolutional networks allowing patient classification and biomarker identification," *Nat. Commun.*, vol. 12, no. 1, p. 3445, Jun. 2021, doi: 10.1038/s41467-021-23774-w.